# An Empirical Evaluation On Vibrotactile Feedback For Wristband System

Feng Wang[1,2], Wanna Zhang[2], and Wei Luo[2]

[1]School of Physics and Electronic Engineering, Guangzhou University, Guangzhou, 510006, China
[2]Key Lab of Computer Technology Application of Yunnan Province, Kunming University of Science and Technology, Kunming 650504, China

Correspondence should be addressed to Feng Wang; fengwang@gzhu.edu.cn

## Abstract

With the rapid development of mobile computing, wearable wrist-worn is becoming more and more popular. But the current vibrotactile feedback patterns of most wrist-worn devices are too simple to enable effective interaction in nonvisual scenarios. In this paper, we propose the wristband system with four vibrating motors placed in different positions in the wristband, providing multiple vibration patterns to transmit multi-semantic information for users in eyes-free scenarios. However, we just applied five vibrotactile patterns in experiments (positional up and down, horizontal diagonal, clockwise circular, and total vibration) after contrastive analyzing nine patterns in a pilot experiment. The two experiments with the same 12 participants perform the same experimental process in lab and outdoors. According to the experimental results, users can effectively distinguish the five patterns both in lab and outside, with approximately 90% accuracy (except clockwise circular vibration of outside experiment), proving these five vibration patterns can be used to output multi-semantic information. The system can be applied to eyes-free interaction scenarios for wrist-worn devices.

## 1. Introduction

In the nowadays information society, wearable devices of different forms have come into people's lives in various aspects, simultaneously posing a challenge to the inter-activity design. With the ability of aiding in the remote monitoring of patients, wearables provide real-time access to health records and provide quicker diagnosis and treatment of conditions. Therefore, more studies began to focus on how to effectively exploit wearable technology in the field of e-Healthy [21] or healthcare system.

In contrast to classical graphical user interfaces, wearable devices cannot provide a good visual interaction experience owing to their diverse designs and size limitation. Auditory feedback is a good way to convey semantic information to the users [19]. However, it is difficult for users to receive audio information effectively in mobile environments, where auditory channels are compromised by external noise and social concerns. At the same time many researches have shown that tactile display, without the drawbacks of visual or auditory display, is an ideal interactive mode for distracting situations [1, 2, 4, 6].

Vibration is the basic tactile feedback pattern of wrist-worn devices [3]. By this means users can be notified without visual load and in private or noisy situations [17]. However, the present vibrotactile displays are too simple, most of them can output only two kinds of information, i.e., "vibrating" and "not vibrating", limiting the feedback information received by users. Thus, users tend to take extra time to confirm what kind of information they receive after sensing vibration. At present, there are many studies to investigate the effects of multiple vibrotactile patterns. But these devices array vibration motors on the same side [8, 11, 18, 20]. Each kind of vibrotactile patterns always generates on the same plane of the wrist skin. The skin area of the wrist is relatively narrow. Users may be confused to confirm which vibration motor is working during they all vibrating.

In order to study the user's identification of multiple vibrotactile patterns through a device spatially arraying vibration motors, we present a wearable wristband prototype which spatially arrays vibration motors around the whole wrist. It can produce many kinds of vibrotactile patterns, allowing users to obtain multi-semantic information transmitted through the tactile channel. As shown in Figure 1, four vibration motors are embedded into a customized wrist-worn device. Motors vibrate in different positions or in different orders can generate various vibrotactile patterns. The wristband prototype we developed could generate nine vibrotactile patterns (four positional patterns, two diagonal patterns, two circular patterns, and one total vibration pattern). Through pre-experiment, we decided to use five patterns to research.

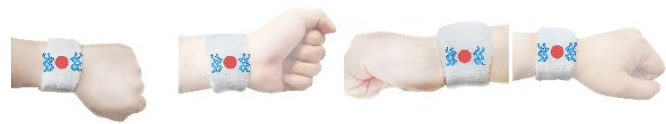

Figure 1: System concept map: vibrators are embedded at four positions in a wristband, corresponding to the middle points of the four sides of the wrist.

In this paper, we first discuss the relevant researches on tactile display for wearable devices, vibrotactile feedback conveying rich messages for wearable devices, and which body locations more sensitive to vibrations feedback. Next, we present our wristband prototype system, and apply five vibrotactile patterns in our study. Then, we describe two experiments that we conducted to examine the accuracy with which users are able to distinguish the five different vibrotactile patterns and the effect of wristband prototype system on the use's ability to perform in lab and outside eyes-free interaction scenarios. Finally, we show some further applications for the wristband prototype system.

## 2. Related Work

Wearable devices have limited visual and auditory information output channels owing to their physical characteristics and complex use scenarios. Tactile display is gaining attentions and has been confirmed as desirable feedback for such devices. Roumen et al. [4] made a comparative study of notification channels (light, vibration, sound, poke, thermal ) for wearable interactive rings. They concluded vibration was the most reliable and fastest channel to convey notification. And Hsieh et al.[5] added tactile feedback in the haptic glove which assists to interact with smart glasses enhancing tangibility. Exploring natural vibrotactile interaction has become the main research direction.

A few researchers have shown different vibrotactile parameters (e.g., intensity, frequency, temporal pattern, spatial pattern) could conveying rich messages for wearable devices. Cauchard et al. [6] encoded the vibrations using the duration and rhythm to represent progress. The ActiVibe they produced utilizing the vibrations have been confirmed with up to 88.7% recognition rate through the outdoors experiment and give a list of factors that affect the recognition rate, such as other vibrations produced during the activity, the materials of device generating uncomfortable feeling and so on. Brewster et al. [7, 8] investigated the perception of Tactons which encode three dimensions of information using three different vibrotactile parameters. The result reveals that spatial patterns are easier to discriminate than frequency and intensity. Van Erp et al. [9] designed a tactile waypoint navigation display consists of eight tactors around the user's waist, and they translated distance to vibration rhythm while the direction was translated into vibration location. Their experiments indicated the usefulness of the tactile display on waypoint navigation. In addition, previous studies also propose richer interaction using several vibrations. Yatani and Truong [10] proposed a real-time feedback system, SemFeel, through multiple vibration motors attached to the backside of a mobile device. This system contains 10 vibration patterns that users can distinguish them at approximately 90% accuracy and it supports accurate eyes-free interactions. Lee and Starner [11] presented a wrist-worn wearable tactile displays that provide easy to perceive alerts for on-the-go users. Their system developed with three actuators in a triangular layout on the volar side of the wrist provides 24 vibration patterns with up to 99% accuracy after 40 minutes of training for users. And the comparison test showed uses' perception of incoming alerts for wrist-worn wearable tactile displays don't decrease when visually distracted. All these works can prove that vibration display is an effective way to improve users interaction experience.

Studies have also been conducted to explore which body locations are more sensitive to vibrations [12, 13]. The result both indicated that wrists are generally better for feeling vibrations relative to other body parts. According to these analysis and results, some wrist systems based on vibrotactile feedback have been proposed. Bosman et al. [17] developed a dual-wrist system to guide a pedestrian inside an unknown building. The vibrations indicated directions and stops. The result suggests vibration tactile output can greatly help improve the performance of this kind of wearable wristband devices and effectively reduce the disruptiveness of technology. Huxtable et al. [14] presented a tactile interface made up of two wristbands that vibrate to signal left and right turns for wayfinding devices designed for cyclists. Dobbelstein et al. [15] presented a bearing-based pedestrian navigation approach that utilizes vibrotactile feedback around the user's wrist to convey information about the general direction of a target. And these mobile prototypes demonstrate their feasibility in the initial navigation research.

In summary, researches on how to use vibrotactile channels to improve the natural interaction of wearable devices and send complex information have achieved good results. But user interface and interactive design of wrist-worn devices have not been studied systematically and thoroughly. Since previous studies reveal that spatial patterns are easy to discriminate [8]. Our research especially focus on how much vibration patterns wrists can accurately distinguish.

## 3. System

To study the acquisition of multi-semantic information sent by wrist-worn devices through vibration in eyes-free inter-active scenarios, we designed a hardware prototype that generates a variety of vibration patterns. The prototype consists of two components (Figure2): a PC and development board (Figure 2-b) to send vibration commands and a wristband device (Figure 2-a) to receive vibration commands and vibrate accordingly. In our experiment, the wristband device need to be wearable by users.

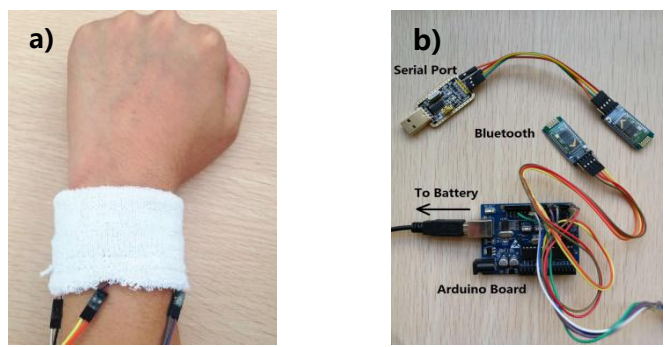

Figure 2: a) front of the wrist-worn device; b) Arduino board connected

*3.1 Sending End*

The physical prototype of the sending end is shown in Figure 2-b. The pattern of the development board that controls the four vibration motors is an Arduino UNO R3. It works as an intermediary to receive signals from a computer by connecting to the serial port of the computer, and it can also control the operating power supplied to each motor by using pulse width modulation (PWM) to change the duty cycle. Therefore, we use this technique to both power on/off motors and to control their vibration intensity.

*3.2 Receiving End*

Considering that we just focus on vibration patterns, a set of basic vibration motors, 12mm in diameter and 3.4mm thick, have been used to output vibrotactile information. Each motor was provided 80mA at 3V in the experiment.

Since wrist-worn devices are in contact with the wrist over a limited area, it is difficult to distinguish many vibration information owing to mutual interference. In addition, referencing the wrist-worn prototype presented by Gupta et al. [16], we selected four locations (see Figure 1) to place four vibration motors, which are connected to an Arduino development board to receive vibration commands received from PC to serial port through bluetooth .

In daily life, there is a certain gap between a wrist-worn device and the wrist, and these small gaps may affect the users' perception of vibrotactile feedback. Therefore, we focus on elastic sports wristbands. As shown in Figure 2-a, this type of wristband is close to the user's wrist during used so that the vibrations of the motors can be accurately perceived. Moreover, since our hardware prototype requires the premise of seamless fitness, the study of vibration patterns for loose wrist-worn devices (such as hand rings) is a subject for future research.

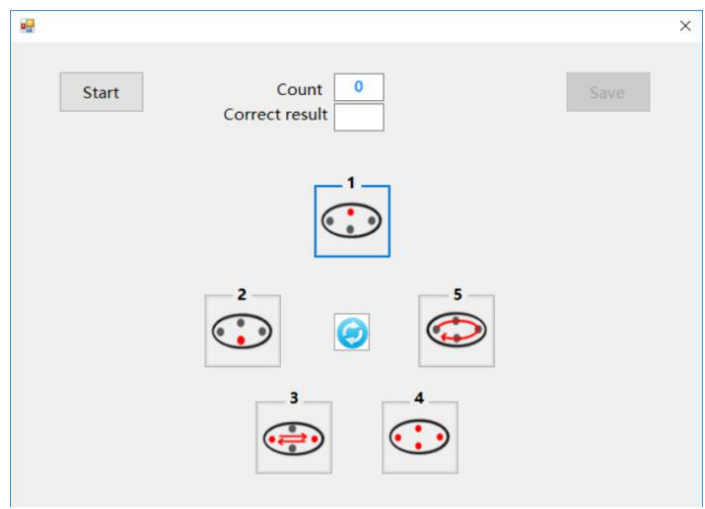

Figure 3: Screenshot of the application running on a Windows machine in the experiments.

### 3.3 Interactive Interface

The running interface of the software program is shown in Figure 4, and the program written with the C# programming language runs on a computer with a Windows 10 operating system. Figure 4 presents the five vibrotactile patterns that we used in the two experiments. There are four types of patterns: positional (up and down), horizontal diagonal, clockwise circular and total. Each icon button of patterns had the same size (90 pixels × 90 pixels). The start button means the tests start and the program starts running. The blue button in the middle requires participants press to randomly generate vibrotactile patterns. The interface also displays the number of test and the result of each test whether right or not. When the counting number reaches the setting number, all the result will be saved automatically.

# 4. Experiment 1：Natural Wristband in lab

## 4.1 Participants and Apparatus

12 participants (4 female, 8 male) from the university participated in the study. They are all students, coming from Kunming University of Science and Technology, who are randomly invited to participate experiments. The age range was 20 to 28 (mean=23.3, SD=1.87). Four participants wore smart wrist devices once previously. The hardware and interactive interface have been introduced in the previous section.

## 4.2 Vibrotactile Feedback Patterns

The only dependent variable that we controled in this experiment is the vibration patterns. As shown in Figure 3, we designed five patterns for the Dancing wristband prototype. These five vibration patterns can be divided into four types: positional vibration (a single motor vibrate once); diagonal vibration (two diagonal motors vibrate four times); circular vibration (four motors vibrate four times sequentially in a ring); and total vibration (all motors vibrate simultaneously once). In addition, the time interval from the start of vibration to the end of vibration was 1 second for all pattern, taking into account fairness for the perceived time of vibration.

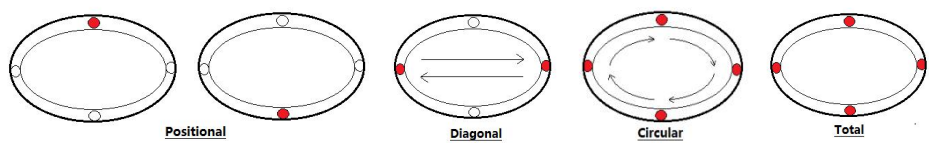

Figure 4: The five vibrotactile feedbanck patterns.

Before deciding on these five patterns, we set up nine patterns in a pilot experiment to explore recognition rates of different vibration patterns. These nine patterns include four positional vibration patterns (up, down, left, right), two diagonal vibration patterns (horizontal and vertical directions), two circular vibration patterns (clockwise and anti-clockwise), and one total vibration pattern. Each vibration pattern repeatedly presented twelve times. The procedure and equipment of the pilot experiment are same as experiment 1. Six participants volunteered for the experiment and the rates of recognition are shown in Table 1.

Table 1: The recognition rate of each pattern.

| Positional | | | | Diagonal | | Circular | | Total |
|---|---|---|---|---|---|---|---|---|
| up | down | left | right | horizontal | vertical | clockwise | anti-clockwise | total |
| 91.7% | 94.4% | 88.9% | 88.9% | 66.7% | 63.9% | 83.3% | 61.1% | 86.1% |

Although only two recognition rates of positional vibration patterns are over 90%, two main problems which should be solved have been found. 1) Vibration patterns of the same type can very easily interfere with each other. For example, participants have difficulty distinguishing horizontal diagonal and vertical vibrotactile patterns. 2) Diagonal vibration patterns and circular vibration patterns have similar vibrotactile feedback and are difficult to distinguish.

To solve these problems, we have taken the following measures. 1) Only one vibration pattern was retained for each type except for the positional vibration type. Through analyzing the results, we found positional vibration were easily to distinguish and two of them were retained. 2) Participants prefer vibration with smoothing over vibration without smoothing [10]. So, we set vibration intensities of the four motors in the circular vibration pattern to be 40%, 60%, 80%, and 100%, respectively. Except for the circular vibration pattern with changing vibration intensity, the vibration intensity was always 100%. This is because we just focused on evaluating how accurately participants can distinguish vibration patterns rather than other vibrotactile parameters.

Finally, we determined the five patterns (positional up and down, horizontal diagonal, clockwise circular, and total vibration) which used in the two experiments. Each of the five patterns had a relatively high recognition rate in there type. There are two advantages. 1) All types were retained to ensure the diversity of vibrotactile feedback. 2) According the "magical number 7, plus or minus two " rule in human-computer interaction, participants could effectively recognize these five patterns in a short time. And we can explore more patterns in future.

The generation order of the five vibration patterns in the experiment was completely random. Each vibration pattern was repeated four times in a block of experiments, and the entire experiment contained four blocks. Hence, each participant took part in:
$$5 \text{ vibration patterns} \times 4 \text{ blocks} \times 4 \text{ repetitions} = 80 \text{ trials in total.}$$

### 4.3 Task

In the experiment, participants were required to feel the vibration patterns randomly generated by the wrist-worn device. As shown in Figure 4, the system randomly generated a vibration pattern after participants clicked the blue button. Participants needed to select a pattern from the five representative vibration patterns according to the vibration they felt, and then clicked the icon button for the selected pattern. Each time after a participant made a choice, the system displayed the correct vibration pattern in the dialog box in black font for a correct choice and red font for an incorrect choice. The number of completed vibrations was displayed in the upper part of the dialog prompt box. Throughout the experiment, participants were asked to complete the task as quickly and accurately as possible.

### 4.4 Procedure

Each participant was informed of the use of this system before the experiment system. And participants needed to wear the wristband part of the system in their non-dominant hand and used their dominant hand to operate the mouse and software program for interaction.

At the same time, participants were asked to wear headphone which was playing music so as not to notice the sound of vibration motors. And the wristband need beyond the vision of participants in order to reproduce an eyes-free situation.

Before the formal experiment started, participants were asked to become familiar with the whole process of the formal experiment and practiced until they adapted to the system. Participants took a short break after each block in the formal experiment. Generally speaking, a participant completed the practice period in about 15 minutes. In total, a complete experiment lasted about 30 minutes.

*4.5 Results and Discussion for the lab study*

We tested two kinds of data: reaction time and error rate. The reaction time contains two parts, the user feeling vibrotactile patterns and choosing patterns, the actual reaction time should be shorter. The error rate for each pattern was calculated per block per participant (i.e., 100 × [the number of the wrong responses] / 4). The results are drawn as bar diagrams showing in Figure 5 and Figure 6. The time variable was measured from the time when motors started to vibrate to the time when participants clicked an icon button to make a choice. After the experiment was completed, the experimental data were collected for calculating statistics and analysis.

Figure 5 shows the average error rates of the five vibration patterns. All the average error rates are below 9%, and the lowest average error rate is positional(down). A one-way analysis of variance (ANOVA) test (the significant level is 0.05) for the average error rates against pattern indicates existence of statistically significant differences (mean = 4.27, SD = 10.73, $F_{4,235}=3.94$, $p<.01$). The post-hoc Turkey multiple comparison revealed that statistically significant differences exist between positional(down) and circular (clockwise) ($p < .05$).

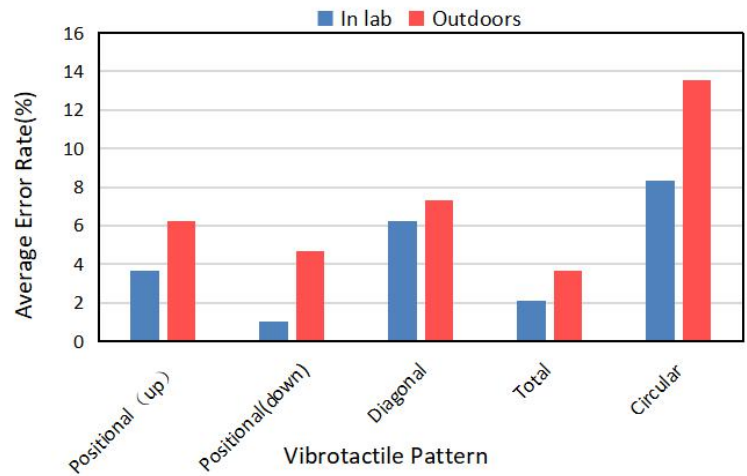

Figure 5: Results of error rate from the laboratory and outside studies.

For the reaction time, all vibrotactile types average reaction time is between 1600-2200(ms). Positional vibrotactile type requires shortest reaction time on average. We made a one-way ANOVA test (the significant level is 0.05) between the four vibration types (positional vibration, diagonal vibration, circular vibration and total vibration). The result shows the existence of statistically significant differences ((mean = 1798.73, SD = 861.18, $F_{3,956}=20.44$, $p<.001$). Then, we combine the two types of positional vibration into one category, so sample sizes is unbalanced. To accommodate the unbalance, a Scheffe multiple comparison was used in the post-hoc test. It indicated that the reaction time for the circular vibration was significantly different from others($p < .001$).

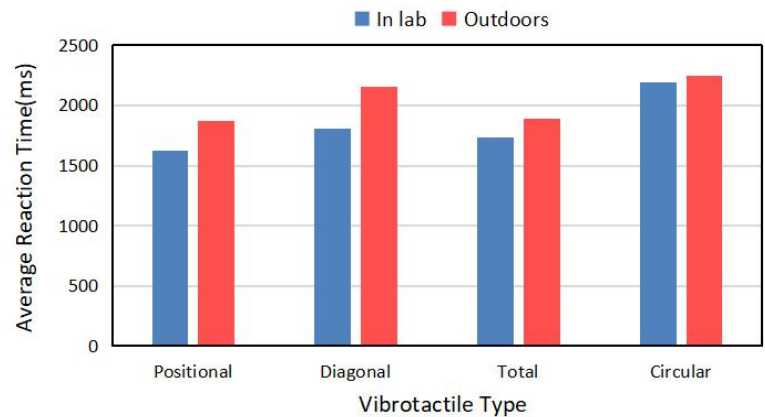

Figure 6: Results of reaction time from the laboratory and outside studies.

Figure 7 shows the average error rates across blocks. Each participant performs an entire experiment containing eighty trials that dividing four blocks. The whole results of all participants are classified according to the four blocks. The average error rates of each block is less than 5%. Block 1 has the lowest average error rate. And the average error rate of blocks show a trend of growth first, then tends to be stable. A one-way ANOVA test (the significant level is 0.05) for the average error rates against Block indicates that there is no statistically significant difference (mean = 4.27, SD = 3.57).

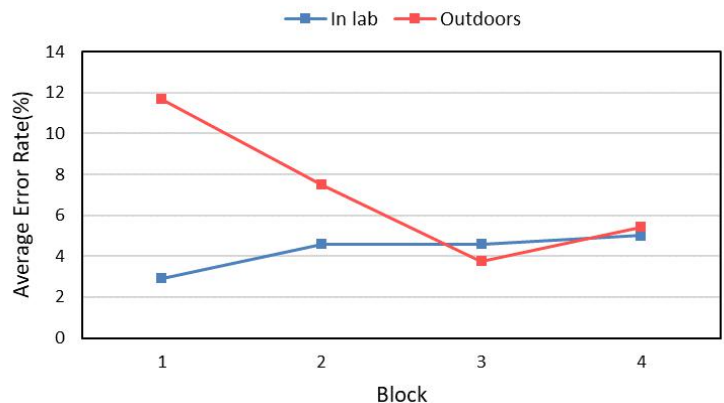

Figure 7: Results of error rate across Blocks from the laboratory and outside studies.

Figure 8 shows the average reaction time across blocks. The average reaction time of each block is between 1700-1900(ms). Block 2 requires the shortest reaction time on average. The trend fluctuates slightly between blocks. A one-way ANOVA test (the significant level is 0.05) for the reaction time against Block indicates that there are no statistically significant difference (mean = 1798.73, SD = 355.40).

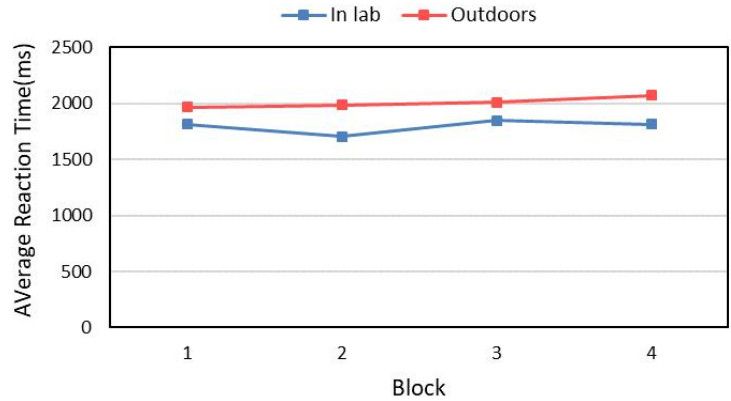

Figure 8: Results of reaction time across Blocks from the laboratory and outside studies .

The results indicates that participants were able to distinguish the five vibrotactile patterns, and recognition accuracy of all patterns are more than 90%. But the recognition accuracy of diagonal pattern and circular pattern is lower than average (average recognition accuracy of all patterns is 95.73% ). And the change of average error rates between blocks, error rate becomes rise between block 1 to block 2. One possible reason is that participants may feel a little numb after long time adapting, Their ability to response stimulus decreases so the reaction time increase. We need to give them enough break time to rest after they adapt the wristband prototype and after finishing each block experiment.

## 5.  Experiment 2：Outdoors Natural Wristband

We used the same equipment to run the experiment and participants of experiment 2 are same as experiment 1. This time participants needed to go outside and not just sitting. They took some activities instead, such as walking, jogging. We need to do various activities in daily life, but the laboratory environment does not represent all the real living environment. We chose the outdoor scene in the university, surrounded by many classrooms and a bus station, where many people would come and go.

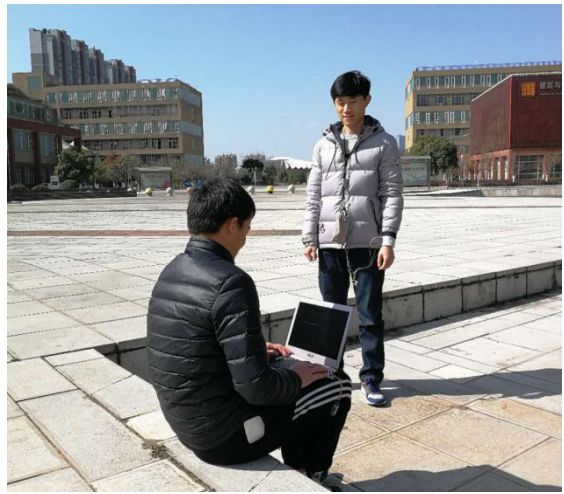

Figure 9: The outdoors experimental setup.

Participants came outdoors to experiment, so an experimental assistant was also required to help the participants choose the vibration types (Figure 9). We also made sure the assistant was within the range of hearing the participants' answers exactly.

*5.1 Procedure*

Participants wore the wristband part of the system in their non-dominant hand to feel vibrotactile patterns. Meanwhile, participants needed to walk within a range rather than standing still. When the participants felt vibration, they were required to speak out which pattern they felt. Then the assistant chose the answer and performed subsequent interactions through the interactive interface.

Before the formal experiment started, participants and the assistant were asked to get familiar with the whole process practiced until they cooperated well. Each participant repeated twenty trials in a block of the formal experiment which consists of four blocks. Participants could take a short break after each block. In total, an entire experiment took about 30 minutes.

*5.2 Results and Discussion for the outdoors study*

Although the number of participants involved in the experiment is limited, some conclusions can be drawn from the statistical analysis. The results are shown in Figures and compared with the results of laboratory study.

Figure 5 shows all the average error rates below 8% except circular vibrotactile pattern (its average error rate is 13.5%). The average error rates is statistically significant different in dynamic state for the five vibration patterns (mean = 7.08, SD = 13.98, $F_{4,235}=3.86$, $p<.01$). And after the post-hoc Turkey multiple comparisons, it discovered that the statistical difference exist between positional(down) pattern and circular pattern , total pattern and circular pattern($p<.05$).

The one-way ANOVA test (the significant level is 0.05) suggests that there is a statistically significant difference between the reaction times under dynamic conditions for the four vibration types (mean = 2009.22, SD = 1162.07, $F_{3,956}=6.29$, $p<.001$) outside. The post-hoc Scheffe multiple comparison revealed that there are statistically significant differences between the two groups: positional pattern and circular pattern ($p<.05$), total pattern and circular pattern ($p<.001$).

The result of error rates across blocks is different from the experiment 1. Its tendency changes from high to low, then stabilizes. A one-way ANOVA test (the significant level is 0.05) for the error rates against blocks indicates the existence of statistically significant differences (mean = 7.08, SD = 6.75, $F_{3,44}=3.59$, $P<.05$). The post-hoc Turkey multiple comparison revealed that statistically significant differences exist between block 1 and block 3 ($p< .05$). However, the one-way ANOVA test results of average reaction time against blocks are the same as the experiment. There is no statistically significant difference between blocks (mean = 2008.28, SD = 438.94).

The results shows that participants can still recognize the five vibrotactile patterns. The recognition accuracy of all patterns are more than 90% except circular pattern (its recognition accuracy is 86.5%). And compared with the results of Experiment 1, all the results of Experiment 2 are higher except the average error rates result of blocks.

Maybe users' perceptual ability will be disturbed when the outdoors is noisy and they are in a dynamic state. Additionally, there is still a gap between the wristband and skin, it will affect participants to feel motors vibrate. These findings prove that users can recognize multiple vibrotactile patterns and the efficacy of our wristband prototype.

## 6. Implications

This research can be applied to nonvisual interaction environments because the vibrotactile channel suffers least from interference originating in external environments compared with other output channels.

- Example 1: navigation for the blind and warning of obstacles in different directions. Walking outdoors is inconvenient for the blind because they cannot visually obtain information, but different vibrotactile feedback patterns on the wrist can provide them with direction navigation and obstacle warnings.

- Example 2: different vibration prompts for different application messages on mobile phones. People in a complex environment often cannot receive mobile phone messages in a timely manner. After establishing contact with a mobile phone via Bluetooth, wristband devices can vibrate accordingly when a mobile phone receives application messages so that users do not miss the messages and can determine the types of messages received.

## 7. Conclusions and Future Work

Existing wristband devices do not support various vibration patterns, thus limiting users' perceptions of information from wristband devices in nonvisual interaction scenarios. To solve this problem, we have developed a vibration feedback-based Dancing wristband system that controls multiple vibration motors embedded in a wristband to generate different vibration patterns for multi-semantic information transfer in eyes-free scenarios. The experiment we conducted verified that users can identify five kinds of vibration patterns, and the experimental results show that users can successfully recognize all patterns at a rate over 90% in the static state, meanwhile, the rate over 85% in the dynamic state. The following factors may affect the users' performance of the system: firstly, the material of the wearable wristband that contains four motors is not thin enough for users to feel vibration; secondly, the contact surfaces of motors (left and right sides of the wrist) and wrist are not large enough, so the users are not sensitive to vibration generating from the two sides.

Although the results in this paper proved the effect of our system, much further research is required. We need to do more further work to enhance users' experience of the wearable wristband according to these possible influencing factors. We can improve users experience through changing the material of wristband or adding other tactile forms to form a mixed tactile pattern. Since the system requires users' to distinguish between different vibration patterns, we will explore more vibration patterns that can be accurately identified. The participants of our experiments are young people, we will explore the availability of this wristband prototype system at different ages.

## 8. acknowledgements

This work is supported by the National Natural Science Foundation (NNSF) of China under Grant 61462053. Part of the manuscript was accepted by the 37th Chinese Control Conference, but the manuscript submitting this time has been significantly rewritten. Sincerely thanks to the project for funding. We appreciate the helpful comments and suggestions of all the reviewers. We also gratefully acknowledge all the participants of our studies for their time and effort.

# References

[1] Brewster, S., Chohan, F., and Brown, L., "Tactile feedback for mobile interactions," in *Proceedings of the SIGCHI Conference on Human Factors in Computing Systems*, pp.159-162, San Jose, California, USA, 2017.

[2] Straughn, S. M., Gray, R., and Tan, H. Z., "To Go or Not to Go: Stimulus-Response Compatibility for Tactile and Auditory Pedestrian Collision Warnings," *IEEE Transactions on Haptics,* vol. 2, no. 2, pp. 111-117, 2009.

[3] Motti, V. G. and Caine, K., "Smart Wearables or Dumb Wearables?: Understanding how Context Impacts the UX in Wrist Worn Interaction," in *Proceedings of the 34th ACM International Conference on the Design of Communication*, pp. 1-10, Silver Spring, MD, USA, 2016.

[4] Roumen, T., Perrault, S. T. and Zhao, S., "NotiRing: A Comparative Study of Notification Channels for Wearable Interactive Rings," in *Proceedings of the 33rd Annual ACM Conference on Human Factors in Computing Systems*, pp. 2497-2500, Seoul, Republic of Korea, 2015.

[5] Hsieh, Y.-T., Jylha, A., Orso, V., Gamberini, L. and Jacucci, G.,"Designing a Willing-to-Use-in-Public Hand Gestural Interaction Technique for Smart Glasses," in *Proceedings of the 2016 CHI Conference on Human Factors in Computing Systems,* pp. 4203-4215,*Santa Clara*, California, USA, 2016.

[6] Cauchard, J. R., Cheng, J. L., Pietrzak, t. and Landy, J. A., "ActiVibe: Design and Evaluation of Vibrations for Progress Monitoring," in *Proceedings of the 2016 CHI Conference on Human Factors in Computing Systems*, pp. 3261-3271,*Santa Clara*, California, USA, 2016.

[7] Brewster, S. and Brown, L. M., "Tactons: structured tactile messages for non-visual information display," in *Proceedings of the fifth conference on Australasian user interface* - Volume 28, pp. 15-23, Dunedin, New Zealand, 2004.

[8] Brown, L. M., Brewster, S. A. and Purchase, H. C., "Multidimensional tactons for non-visual information presentation in mobile devices," in *Proceedings of the 8th conference on Human-computer interaction with mobile devices and services*, pp. 231-238, Helsinki, Finland, 2006.

[9] Van Erp, J. B. F., Van Veen, H. A. H. C. and Jansen, C., "Waypoint navigation with a vibrotactile waist belt," *ACM Trans.* Appl. Percept., vol. 2, no. 2, pp. 106-117, 2005.

[10] Yatani, K. and Truong, K. N., "SemFeel: a user interface with semantic tactile feedback for mobile touch-screen devices," in *Proceedings of the 22nd annual ACM symposium on User interface software and technology*, pp. 111-120, Victoria, BC, Canada, 2009.

[11] Lee, S. C. and Starner, T., "BuzzWear: alert perception in wearable tactile displays on the wrist," in *Proceedings of the SIGCHI Conference on Human Factors in Computing Systems, Atlanta*, pp. 433-442, Georgia, USA, 2010.

[12] Karuei, I., MacLean, K. E., Foley-Fisher, Z., MacKenzie, R., Koch, S. and EI-Zohairy, M., "Detecting vibrations across the body in mobile contexts," in *Proceedings of the SIGCHI Conference on Human Factors in Computing Systems*, pp. 3267-3276, Vancouver, BC, Canada, 2011.

[13] Machida, T., Dim, N. K. and Ren, X., "Suitable Body Parts for Vibration Feedback in Walking Navigation Systems," *International Symposium of Chinese Chi,* pp. 32-36, 2015.

[14] Huxtable, B. J., Lai, K. H., Zhu, J. W. J., Lam, M. Y. and Choi, Y. T., "Ziklo: bicycle navigation through tactile feedback," in *CHI '14 Extended Abstracts on Human Factors in Computing Systems*, Toronto, Ontario, Canada, pp. 177-178, Toronto, Ontario, Canada, 2014.

[15] Dobbelstein, D., Henzler, P. and Rukzio, E., "Unconstrained Pedestrian Navigation based on Vibro-tactile Feedback around the Wristband of a Smartwatch," in *Proceedings of the 2016 CHI*

*Conference Extended Abstracts on Human Factors in Computing Systems*, pp. 2439-2445, Santa Clara, California, USA, 2016.

[16] Gupta, A., Pietrzak, T., Roussel, N. and Balakrishnan, R., "Direct Manipulation in Tactile Displays," in *Proceedings of the 2016 CHI Conference on Human Factors in Computing Systems*, pp. 3683-3693, Santa Clara, California, USA, 2016.

[17] Bosman, S., Groenendaal, B., Findlater, J. W., Visser, T., Graaf, M. and Markopoulos, P., "GentleGuide: an exploration of haptic output for indoors pedestrian guidance," *Human-computer Interaction with Mobile Devices & Services*, pp. 358-362, 2003.

[18] Wang, Y., Millet, B. and Smith, J. L., "Designing wearable vibrotactile notifications for information communication," *International Journal of Human-Computer Studies,* Volume 89 Issue C, pp. 24-34, 2016.

[19] Zhao, S., Dragicevic, P., Chignell, M., Balakrishnan, R. and Baudisch, P., "earPod: Eyes-free Menu Selection using Touch Input and Reactive Audio Feedback," in *Proceedings of the 2007 CHI Conference on Human Factors in Computing Systems,* pp. 2439-2445, San Jose, California, USA, 2007.

[20] Lee, J., Han, J. and Lee, G., "Investigating the Information Transfer Efficiency of a 3x3 Watch-back Tactile Display," in *Proceedings of the 2015 CHI of the 33rd Annual ACM Conference on Human Factors in Computing Systems,* pp. 1229-1232, Seoul, Republic of Korea, 2015.

[21] González-Landero, F., García-Magariño, I., Lacuesta, R. and Lloret, J., "Green Communication for Tracking Heart Rate with Smartbands," *Sensors (Basel),* vol. 18(8), pp. 2652, 2018.